# PATH INTERGRAL APPROACH TO THE KINEMATICAL BROWNIAN MOTION, DUE TO A RANDOM CANONICAL TRANSFORMATION.

By

Tchoffo M. [†,+], Beilinson A. A. [‡]

[†] Department of Physics, Faculty of Science, University of Dschang, Cameroon

[‡] Department of Theoretical Physics, Russian Peoples Friendship University, Russia.

**ABSTRACT**

The **stochastization** of the Jacobi second equality of classical mechanics, by Gaussian white noises for the Lagrangian of a particle in an arbitrary field is considered. The quantum mechanical Hamilton operator similar to that in Euclidian quantum theory is obtained. The conditional transition probability density of the presence of a Brownian particle is obtained with the help of the functional integral. The technique of factorisation of the solution of Fokker – Plank equation is employed to evaluate the effective potential energy.

---

[+] corresponding author : e-mail : mtchoffo2000@yahoo.fr

# 1 Introduction

This paper investigates stochastic equations that describe random canonical transformations of classical Hamilton systems and the relationship with Euclidean quantum theory. We obtain a stochastic equation that corresponds to a fundamental state function and is a positive definite solution of the corresponding Fokker–Planck equation ([1-3]

In Euclidean quantum theory every Hamilton operator correspond to an entire set of stochastic equations. Each of the equations result from the solution of the time reversed Bloch equation with the given Hamiltonian [1,2,4]. For the Hamilton operator $\hat{H}$, the corresponding stochastic equation has the form:

$$\dot{x} - \frac{1}{2}\frac{\partial}{\partial x}\ln \widetilde{Z} = \dot{\varphi}$$

where $\varphi(t)$ is a Wiener process [5], and $\widetilde{Z}$ satisfies the time reversal Bloch equation

$$\frac{\partial \widetilde{Z}}{\partial t} = \hat{H}\widetilde{Z} \quad , \quad \hat{H} = -\frac{1}{4}\frac{\partial^2}{\partial x^2} + V(x,t) \qquad 1.1$$

The solution of time reversal Bloch equation 1.1 yields the corresponding Fokker-Plank equation:

$$\frac{\partial W}{\partial \tau} + \frac{1}{2}\frac{\partial}{\partial x}\left(\frac{\partial}{\partial x}\ln \widetilde{Z}(x,t)W\right) = \frac{1}{4}\frac{\partial^2 W}{\partial x^2} \qquad 1.2$$

In order to solve 1.2, consider the direct Bloch equation

$$\frac{\partial Z}{\partial t} = -\hat{H}Z$$

with the fundamental solution $Z(x_0,0;x,t)$ written through the Feynman-Kăc formula[1,2,6]:

$$Z(x_0,0;x,t) = \int_c \exp\left[-\int_0^t V[x(\tau)]d\tau\right] d_w x(\tau) \quad , \quad x(0) = x_0 \quad , \quad x(t) = x$$

where

$$d_w x(\tau) = \exp\left[-\int_0^t \dot{x}^2(\tau)d\tau\right]\prod_{\tau=0}^t \frac{dx(\tau)}{\sqrt{\pi d\tau}}$$

is the Wiener measure [5]. Consequently the fundamental solution of 1.2 considering the Wiener integral [4] yields

$$W(x_0,0;x,t) = \int_C \exp\left[-\int_0^t \left\{\dot{x} - \frac{1}{2}\frac{\partial}{\partial x}\ln\tilde{Z}^2 - \frac{1}{4}\frac{\partial^2}{\partial x^2}\ln\tilde{Z}\right\}d\tau\right]\prod_{\tau=0}^t \frac{dx_\tau}{\sqrt{\pi d\tau}} \qquad 1.3$$

$$x(0) = x_0, \; x(t) = x, \; x(\tau) = x_\tau$$

where

$$\exp\left[-\frac{1}{4}\int_0^t \frac{\partial^2}{\partial x^2}\ln\tilde{Z}(x,\tau)d\tau\right] \qquad 1.4$$

plays the role of the transformation Jacobian of the following Fredholm integral equation (Volterra-type and equivalent to the initial stochastic equation):

$$x(t) - \frac{1}{2}\int_0^t \frac{\partial}{\partial x}\ln\tilde{Z}(x,\tau)d\tau = \varphi(t) \qquad 1.5$$

It follows that 1.4 is the Fredholm denominator for 1.5. From 1.3 and considering

$$\frac{d}{d\tau}\ln\tilde{Z}(x,\tau) = \frac{\partial \ln\tilde{Z}(x,\tau)}{\partial x\tau}\dot{x}_\tau + \frac{\partial \ln\tilde{Z}(x,\tau)}{\partial \tau}$$

we arrive at

$$W(x_0,0;x,t) = \frac{\tilde{Z}(x,\tau)}{\tilde{Z}(x_0,0)}Z(x_0,0;x,t)$$

that is the so-called factorisation theorem of the solution of Fokker – Planck equation [1,2,4,7] and has the following properties:

1. $\int_{-\infty}^{+\infty} W(x,0;x,t)dx = 1$      1.6

2. $\lim_{x \to x_0} W(x_0,0;x,t) = \delta(x-x_0)$      1.7

3. $\int_{-\infty}^{+\infty} W(x_0,0;x_\tau,\tau)W(w_\tau,\tau;x,t)dx_\tau = W(x_0,0;x,t)$      1.8

for any $\tau$, $0 \leq \tau \leq t$.

Property 1.8 is the Einstein – Smolukhovsky – Kolmogorov equation and processes for which it is satisfied are Markovian.

## 2 FREE PARTICLE

Consider the stochastic equation

$$\dot{x}_\tau + \frac{1}{m} \frac{\partial S_{cl}(x_\tau, \tau; x, t)}{\partial x_\tau} = \dot{\varphi}(\tau) \qquad 2.1$$

that is the perturbed Jacobi second equalities [8]. Here

$$S_{cl}(x_\tau, \tau; x, t) = \int_\tau^t L(x_s, \dot{x}_s, s) ds$$

is the action functional along the classical trajectory and $\varphi(\tau)$ is the Wiener process.

Consider now the following Wiener measure:

$$d_w \varphi(\tau) = \exp\left[-\frac{m}{2\hbar} \int_0^\tau \dot{\varphi}^2(s) ds\right] \prod_{s=0}^\tau \frac{d\varphi(s)}{\sqrt{\frac{2\pi\hbar ds}{m}}}$$

Here $\hbar$ has the sense of the intensity of Wiener process $\varphi(t)$.

In further evaluations we consider the random perturbation of Jacobi second equalities since for the first equalities the Fredholm denominator is degenerate. Consider a free particle action functional

$$S_{cl}^{(0)}(x_\tau, \tau; x, t) = \frac{m}{2} \frac{(x_t - x_\tau)^2}{t - \tau}$$

then 2.1 yields

$$\dot{x}_\tau - \frac{x_t - x_\tau}{t - \tau} = \dot{\varphi}(\tau) \qquad 2.2$$

From here the corresponding Fokker – Planck equation we have:

$$\hbar \frac{\partial W}{\partial \tau} + \hbar \frac{\partial}{\partial x_\tau}\left(\frac{x_t - x_\tau}{t - \tau} W\right) = \frac{\hbar^2}{2m} \frac{\partial^2 W}{\partial x_\tau^2}$$

With the help of the factorisation theorem of the solution of Fokker-Planck equation [1,2], its fundamental solution has the form

$$W(x_0,0;x_\tau,\tau;x,t) = \frac{\exp\left[-\frac{m}{2\hbar}\frac{(x_\tau - x_0)^2}{\tau} - \frac{m}{2\hbar}\frac{(x - x_\tau)^2}{t-\tau} + \frac{m}{2\hbar}\frac{(x - x_0)^2}{t}\right]}{\sqrt{\frac{2\pi\hbar\tau(t-\tau)}{mt}}} =$$

$$= \frac{\exp\left[-\frac{m}{2\hbar}\frac{t}{\tau(t-\tau)}\left(x_\tau - \frac{x-x_0}{t}\tau - x_0\right)^2\right]}{\sqrt{\frac{2\pi\hbar\tau(t-\tau)}{mt}}}$$

2.3

that is positive definite when $0 \leq \tau \leq t$ and can be interpreted as a transition probability density. Thus 2.3 can safely be written precisely through path integral, in the Wiener integral form. It follows that $W(x_0,0;x_\tau,\tau;x,t)$ represents the probability density that describes the presence of Brownian particle at point $x_\tau$ at the moment $t$. This is on condition, that at the initial moment $\tau = 0$ the particle is found at point $x_0$ and at the subsequent moment $t$ at point $x$. Partition the time interval $[0,\tau]$ by the points $\tau_1, \tau_2, \cdots, \tau_j, \cdots, \tau_n = \tau$ the probabilities of passage of the particles through $(n-1)$ doors has the form:

$$\int_{a_1}^{b_1} dx_1 \ldots \int_{a_{n-1}}^{b_{n-1}} dx_{n-1} \prod_{j=1}^{n} \frac{\exp\left[-\frac{m}{2\hbar}\frac{t-\tau_{j-1}}{\Delta\tau_j(t-\tau_j)}\left(x_j - x_{j-1} - \frac{x - x_{j-1}}{t-\tau_j}\Delta\tau_j\right)^2\right]}{\sqrt{\frac{2\pi\hbar\Delta\tau_j(t-\tau_{j-1})}{m(t-\tau_j)}}}, \quad \Delta\tau_j = \tau_j - \tau_{j-1}$$

2.4

and called probabilistic measure of trajectories, satisfying the conditions.

$$a_j \leq x \leq b_j, \quad j = 1, \cdots, (n-1)$$

Reduce size of doors in 2.4 such that $\max_{1\le j\le n} \Delta\tau_j = 0$ then we arrive a single trajectory that passes through a fine tube. This is on condition that at the moment $t$ it passes through point $x$. The probability 2.4 now takes the form:

$$\lim_{\substack{\max \Delta\tau_j \to 0 \\ 1\le j\le n}} \exp\left[-\sum_{j=1}^{n} \frac{m}{2\hbar} \frac{t-\tau_{j-1}}{t-\tau_j} \left(\frac{x_j - x_{j-1}}{\Delta\tau_j} - \frac{x - x_j}{t-\tau_j}\right)^2 \Delta\tau_j\right] \prod_{j=1}^{n} \frac{dx_j}{\sqrt{\frac{2\pi\hbar\Delta\tau_j}{m}}\left(1 - \frac{\Delta\tau_j}{t-\tau_j}\right)} =$$

$$= \frac{1}{\sqrt{(t-\tau)t^{-1}}} \exp\left[-\int_0^\tau \frac{m}{2\hbar}\left(\dot{x}_s - \frac{x-x_0}{t-s}\right)^2 ds\right] \prod_{s=0}^{\tau} \frac{dx_s}{\sqrt{\frac{2\pi\hbar}{m}ds}}$$

Thus

$$W(x_0,0; x_\tau,\tau; x,t) = \frac{1}{\sqrt{(t-\tau)t^{-1}}} \int_C \exp\left[-\int_0^\tau \frac{m}{2\hbar}\left(\dot{x}_s - \frac{x-x_s}{t-s}\right)^2 ds\right] \prod_{s=0}^{\tau} \frac{dx_s}{\sqrt{\frac{2\pi\hbar}{m}ds}} \qquad 2.5$$

This measure could also be obtained directly from 2.2

**3 HARMONIC OSCILLATOR**

From classical mechanics, the Lagrangian of a harmonic oscillator is:

$$L(x,\dot{x},t) = \frac{m}{2}\left(\dot{x}^2 - \omega^2 x^2\right)$$

and the eikonal is:

$$S_{cl}^{(\omega)} = \frac{m}{2} \frac{\omega}{\sin\omega(t-\tau)}\left[(x_\tau^2 + x^2)\cos\omega(t-\tau) - x_\tau, x\right] \qquad 3.1$$

Consider the stochastic equation

$$\dot{x}_t + \frac{1}{m}\frac{\partial S_{cl}^{(\omega)}(x_\tau,\tau;x,t)}{\partial x_t} = \dot{\varphi}(\tau) \qquad 3.2$$

Substitute 3.1 in 3.2 then this yields

$$\dot{x}_\tau - \omega \frac{x - x_\tau \cos\omega(t-\tau)}{\sin\omega(t-\tau)} = \dot{\varphi}(\tau)$$

and its corresponding Fokker-Planck equation is :

$$\hbar \frac{\partial W}{\partial \tau} + \omega \hbar \frac{\partial}{\partial x_\tau}\left( \frac{x - x_\tau \cos\omega(t-\tau)}{\sin\omega(t-\tau)} W \right) = \frac{\hbar^2}{2m} \frac{\partial^2 W}{\partial x_\tau^2}$$

and its fundamental solution is

$$W(x_0,0; x_\tau,\tau; x,t) =$$

$$= \frac{\exp\left[ -\frac{m\omega\cos\omega t}{2\hbar \sin\omega\tau \sin\omega(t-\tau)}\left( x_\tau - \frac{x_\tau - x_0 \cos\omega t}{\sin\omega t}\sin\omega\tau - x_0 \cos\omega\tau \right)^2 \right]}{\sqrt{\frac{2\hbar\pi \sin\omega\tau \sin\omega(t-\tau)}{m\omega \sin\omega t}}}$$

3.3

that is positive definite when $0 \leq \tau \leq t$ and can be interpreted as the probability density. Partition in 3.3 the time interval $[0,\tau]$ by the points $\tau_1, \tau_2, \cdots, \tau_j, \cdots, \tau_n = \tau$ then the probabilities of passage of the particles through $(n-1)$ doors has the form:

$$\int_{a_1}^{b_1} dx_1 \cdots \int_{a_{n-1}}^{b_{n-1}} dx_{n-1} \prod_{j=1}^{n} \times$$

$$\times \frac{\exp\left[ -\frac{m\omega\cos\omega(t-\tau_{j-1})}{2\hbar \sin\omega\Delta\tau_j \sin\omega(t-\tau_j)}\left( x_j - x_{j-1}\cos\omega\Delta\tau_j - \frac{x_j - x_{j-1}\cos\omega(t-\tau_{j-1})}{\sin\omega(t-\tau_{j-1})} \right)^2 \right]}{\sqrt{\frac{2\hbar\pi \sin\omega\Delta\tau_j \sin\omega(t-\tau_j)}{m\omega \sin\omega(t-\tau_{j-1})}}}$$

3.4

Expand the sine and cosine functions in a Taylor series relative to $\tau_j$ and considering $\max_{1 \leq j \leq n} \Delta\tau_j = 0$. 3.4 results at the expression

$$\lim_{\substack{\max \Delta \tau_j \to 0 \\ 1 \le j \le n}} \exp\left[ -\sum_{j=1}^{n} \frac{m}{2\hbar} \frac{t-\tau_{j-1}}{t-\tau_j} \left( \frac{x_j - x_{j-1}}{\Delta \tau_j} - \omega \frac{x_j - x_{j-1}}{\sin\omega(t-\tau_{j-1})} \cos\omega(t-\tau_{j-1}) \right)^2 \right] \times$$

$$\times \prod_{j=1}^{n} \frac{\Delta x_j}{\sqrt{\frac{2\hbar\pi \sin\omega\Delta\tau_j \sin\omega(t-\tau_j)}{m\omega\sin\omega(t-\tau_{j-1})}\Delta\tau_j}} =$$

$$= \sqrt{\frac{\sin\omega t}{\sin\omega t(t-\tau)}} \exp\left[ -\frac{m}{2\hbar} \int_0^\tau \left( \dot{x}_s - \omega \frac{x - x_s \cos\omega(t-s)}{\sin\omega(t-s)} \right)^2 ds \right] \prod_{s=0}^{\tau} \frac{dx_s}{\sqrt{\frac{2\pi\hbar}{m}ds}}$$

$$W(x_0,0;x_\tau,\tau;x,t) = \sqrt{\frac{\sin\omega t}{\sin\omega t(t-\tau)}} \exp\left[ -\frac{m}{2\hbar} \int_0^\tau \left( \dot{x}_s - \omega \frac{x - x_s \cos\omega(t-s)}{\sin\omega(t-s)} \right)^2 ds \right] \times$$

$$\times \prod_{s=0}^{\tau} \frac{dx_s}{\sqrt{\frac{2\pi\hbar}{m}ds}}$$

3.5

that can also be obtained directly form 3.2. Compare 2.5 and 3.5 then we have

$$x_\tau - \int_0^\tau \frac{x - x_s}{t - s} ds = x_\tau - \int_0^\tau \frac{x - x_s \cos\omega(t-s)}{\sin\omega(t-s)} ds = \varphi(\tau) \qquad 3.6$$

It follows from here that the transformation Jacobian is nondegenerated:

$$J = \sqrt{\frac{t-\tau}{t}} \sqrt{\frac{\sin\omega t}{\sin\omega(t-\tau)}} \ne 0$$

and in particular

$$\lim_{\tau \to t} J = \sqrt{\frac{\sin\omega t}{\omega t}} \ne 0$$

From 3.2 and 3.6 we arrive at the equality

$$x_\tau - \int_0^\tau \frac{x - x_s}{t - \tau} ds = x_\tau - \omega \int_0^\tau \frac{x - x_s \cos\omega(t-s)}{\sin\omega(t-s)} ds = x_\tau + \frac{1}{m} \int_0^\tau \frac{\partial S_{cl}(x_s, s; x, t)}{\partial x_s} ds = \varphi_\tau$$

This implies that the non-degeneracy of the Jacobian may permit us to make transformations from Fredholm integrals for a free particle to Fredholm integrals for a particle in arbitrary fields and vice versa.

**4 PARTICLE IN AN ARBITRARY FIELD**

We consider now $W(x_0,0;x_\tau,\tau;x,t)$ for a particle in an arbitrary field. Consider the stochastic equation

$$\dot{x}_\tau + \frac{1}{m}\frac{\partial S_{cl}(x_s,\tau;x,t)}{\partial x_\tau} = \dot{\varphi}_\tau \qquad 4.1$$

$S_{cl}(x_\tau,\tau;x,t)$ Satisfies the reversal Hamilton – Jacobi's equation [8]:

$$\frac{\partial S_{cl}(x_\tau,\tau;x,t)}{\partial \tau} + \frac{1}{2m}\left(\frac{\partial S_{cl}(x_\tau,\tau;x,t)}{\partial x}\right)^2 - V[x(\tau)] = 0 \qquad 4.2$$

The Newton – Lagrange dynamical equation, corresponding to 4.1:

$$m\ddot{x}_\tau = -\frac{\partial V(x_\tau)}{\partial x_\tau} + m\left[\ddot{\varphi} - \frac{\partial}{\partial x_\tau}(\dot{x}_\tau \varphi(\tau))\right] \qquad 4.3$$

It is the force dependent on the state of the system and may not be interpreted as the exterior "cause" of the Brownian motion for a particle in an arbitrary field. Consequently the Brownian motion has no dynamical, but kinematical nature.

The Fokker-Planck equation for the stochastic equation 4.1 has the form

$$\hbar\frac{\partial W}{\partial \tau} - \frac{\hbar}{m}\frac{\partial}{\partial x_\tau}\left(\frac{\partial S_{cl}(x_\tau,\tau;x,t)}{\partial x_\tau}W\right) = \frac{\hbar^2}{2m}\frac{\partial^2 W}{\partial x_\tau^2}$$

The fundamental solution of this equation has the form

$$W(x_0,0;x_\tau,\tau;x,t) = \int_C \exp\left[\int_0^\tau \left(-\frac{m}{2\hbar}\left(\dot{x}_\tau + \frac{1}{m}\frac{\partial S_{cl}(x_s,s;x,t)}{\partial x_\tau}\right)^2 + \frac{1}{2m}\frac{\partial S_{cl}(x_s,s;x,t)}{\partial x_\tau}\right)ds\right]\prod_{s=0}^{\tau}\frac{dx_s}{\sqrt{\frac{2\pi\hbar}{m}ds}}$$

4.4

Consider the total derivative

$$\frac{dS_{cl}(x_\tau,\tau;x,t)}{d\tau} = \frac{\partial S_{cl}(x_\tau,\tau;x,t)}{\partial x}\dot{x}_\tau + \frac{\partial S_{cl}(x_\tau,\tau;x,t)}{\partial x}$$

then from 4.2 and 4.4 we arrive at the expression:

$$W(x_0,0;x_\tau,\tau;x,t) = \frac{\exp\left[-\frac{1}{\hbar}S_{cl}(x_\tau,\tau;x,t)\right]}{\exp\left[-\frac{1}{\hbar}S_{cl}(x_0,0;x,t)\right]} Z(x_0,0;x_\tau,\tau) \qquad 4.5$$

where

$$Z(x_0,0;x_\tau,\tau) = \int_C \exp\left[-\int_0^\tau \left(V(x(s),s) - \frac{1}{2m}\frac{\partial^2 S_{cl}(x(s),s;x,t)}{\partial x_s^2}\right)ds\right]d_w x_s$$

is Kăc's formula[1,2].

In addition to properties (1.11-1.13), the function $W(x_0,0;x_\tau,\tau;x,t)$ in 4.5 also has the following properties:

1 $\lim_{\hbar\to 0} W(x_0,0;x_\tau,\tau,x,t) = \delta(x_0 - x_\tau)dx_\tau$

where $x_\tau$ is the extremum of the variation of the action functional.

2 $\lim_{\substack{\hbar\to 0\\ \tau\to 0}} W(x_0,0;x_\tau,\tau,x,t) = \delta\left(\dot{x}_0 + \frac{1}{m}\frac{\partial S_{cl}(x_0,0;x,t)}{\partial x_0}\right)d\ddot{x}_0$

3. $\lim_{\substack{\frac{\hbar}{\tau(t-\tau)}\to 0\\ t\to 0}} W(x_0,0;x_\tau,\tau,x,t)dx_\tau = \delta\left(\ddot{x}_0 + \frac{1}{m}\frac{\partial V(x_0)}{\partial x_0}\right)d\ddot{x}_0$

where $V$ is the potential energy.

We can safely find $\hat{H}$ from

$$\hbar\frac{\partial Z(x_0,0;x,\tau)}{\partial \tau} = -\hat{H}Z(x_0,0;x_\tau,\tau)$$

and

$$\frac{\hbar \partial}{\partial \tau} \exp\left[-\frac{1}{\hbar} S_{cl}(x_\tau,\tau;x,t)\right] = \hat{H} \exp\left[-\frac{1}{\hbar} S_{cl}(x_\tau,\tau;x,t)\right]$$

where

$$\hat{H} = -\frac{\hbar^2}{2m}\frac{\partial^2}{\partial x_\tau^2} + \left(V(x_\tau,\tau) - \frac{\hbar}{2m}\frac{\partial^2 S_{cl}(x_\tau,\tau;x,t)}{\partial x_s^2}\right)$$

is the Hamilton operator with an effective potential energy:

$$V_{eff} = V(x_\tau,\tau) - \frac{\hbar}{2m}\frac{\partial^2 S_{cl}(x_\tau,\tau;x,t)}{\partial x_s^2}$$

Write equation 4.3 in the form

$$\dot{x}_\tau + \frac{1}{m}\frac{\partial(S_{cl}(x_\tau,\tau;x,t) - mx_\tau\dot{\varphi})}{\partial x_\tau} = 0$$

that is interpreted as a random canonical transformation with the random productive function $x(\tau)\dot{\varphi}(\tau)$. For this reason it is impossible to gauge the canonical transformation.

Coordinates are invariant for the random canonical transformation, and the momentum changes from $P$ to $P' = P - m\varphi$ and thus the transformation is "random world quake"(i.e., the right hand side in equation 4.1 has the same form for all particles).

**CONCLUSION**

We present a path integral approach to kinematical Brownian motion, due to random canonical transformation. This yields the Jacobi's second equalities in the form of a stochastic equation that leads to the corresponding Fokker – Planck equation. Using the theorem of the factorisation of the solution of the Fokker – Planck equation we obtained the Hamilton's operator in which appears an effective potential energy that resembles the Madelung's potential energy. The solution of the Fokker – Planck equation yields the conditional transition probability density $W(x_o,0;x_\tau,\tau;x,t)$ that has additional properties to those in [1,2,4,7]. It is found that the variant of Brownian motion has a direct relationship with the problems of Euclidian quantum mechanics.

**ACKNOWLEDGMENTS**

The authors would like to thank professor Miskievich of the University of Guadalajara (Mexico) and Dr. Fai Cornelius for their invaluable assistance.